\documentclass[12pt,a4paper]{article}

\usepackage{amsmath}
\usepackage{amssymb}
\usepackage{graphicx}
\usepackage{dsfont}
\usepackage{overpic}
\usepackage{cite}

\let\oldappendix=\appendix
\let\oldsection=\section
\renewcommand{\appendix}{\oldappendix%
\def\theequation{\Alph{section}.\arabic{equation}}%
\renewcommand{\section}{\setcounter{equation}{0}\oldsection}}

\newcommand{\beq}{\begin{equation}}
\newcommand{\eeq}{\end{equation}}
\newcommand{\beqa}{\begin{eqnarray}}
\newcommand{\eeqa}{\end{eqnarray}}
\newcommand{\no}{\nonumber}
\newcommand{\q}{\quad}
\newcommand{\qq}{\qquad}

\newcommand{\tr}{\mbox{tr}}

\newcommand{\newop}[2]{\def#1{\mathop{\mathrm{#2}}\nolimits}}
\newop{\artanh}{artanh}
\newop{\det}{det}
\newop{\tr}{tr}
\newop{\diag}{diag}
\newop{\Re}{Re}
\newop{\Im}{Im}

\begin{document}

\hfill 

\hfill 

\bigskip\bigskip

\begin{center}

{{\Large\bf  Renormalization of two-loop diagrams\\[0.3cm] in scalar lattice field theory}}

\end{center}

\vspace{.4in}

\begin{center}
{\large B.~Borasoy\footnote{email: borasoy@itkp.uni-bonn.de} and
        H.~Krebs\footnote{email: hkrebs@itkp.uni-bonn.de}}

\bigskip

\bigskip

Helmholtz-Institut f\"ur Strahlen- und Kernphysik (Theorie), \\
Universit\"at Bonn, Nu{\ss}allee 14-16, D-53115 Bonn, Germany \\

\vspace{.2in}

\end{center}

\vspace{.7in}

\thispagestyle{empty} 

\begin{abstract}
We present a method to calculate to very high precision the coefficients of the divergences
occuring in two-loop diagrams for a massive scalar field on the lattice.
The approach is based on coordinate space techniques and extensive use
of the precisely known Green's function. 
\end{abstract}
\bigskip
\bigskip
\bigskip
\bigskip




\vfill


\section{Introduction} \label{sec:intro}

Lattice regularization is a convenient tool to study both the perturbative and
non-perturbative aspects of the underlying theory. Within this framework the action
is formulated on a discretized space-time lattice and the partition function can be
evaluated numerically, e.g., by employing Monte Carlo simulations. This feature
of lattice regularization allows one to study the non-perturbative regime of a theory 
and is thus of particular interest for QCD at low energies.

Eventually, however, one is interested in performing the continuum limit by letting
the lattice spacing approach zero. In order to establish the correct link between
numerical lattice simulations and the physical continuum, perturbative lattice
calculations prove useful. They are, e.g., necessary to calculate the renormalization
of the couplings in the Lagrangian or the renormalization factors of operator
matrix elements. In general, it allows one to study perturbatively how simulation
results approach continuum results. The discretization effects, i.e., the corrections
due to the finite lattice spacing, can also be investigated and this knowledge may
help to reduce the pertinent systematic error in lattice results.

The explicit calculation of loop diagrams in lattice regularization turns out to
be more tedious than in conventional continuum regularization schemes such as
dimensional regularization.
In this respect, coordinate space methods have proven useful in the evaluation
of Feynman diagrams and allow a very precise determination of the continuum limit of one- and
two-loop integrals \cite{LW}. In \cite{LW} this technique was applied to massless propagators.
In \cite{BK} the formalism of \cite{LW} was extended to massive scalar fields 
by presenting an efficient method to calculate the associated Green's function
to very high precision.

In the present investigation, we will show that the precise knowledge of the
Green's function is helpful in extracting the divergent pieces of two-loop
diagrams for massive scalar fields to very high accuracy. The method developed
is therefore suited to calculate the renormalization of the bare Lagrangian parameters
at the two-loop level.

This work is organized as follows. First, external momenta in the loop propagators
are eliminated by applying the BPHZ scheme on the lattice so that the discussion
can be reduced to sunset-type loops without external momenta in the propagators.
We illustrate the BPHZ method in the next section for the basic sunset diagram.
Then, the divergences of the reduced integrals are evaluated in Sec.~\ref{sec:coord} 
by applying coordinate space methods. In Sec.~\ref{sec:numerics} we present
the numerical results for the quadratic and logarithmic
would-be divergences of the sunset diagram and in Sec.~\ref{sec:relations} we
provide additional relations between the logarithmic pieces and one-loop tadpole diagrams.
We summarize our findings in Sec.~\ref{sec:concl}.
Technicalities are deferred to the appendices.

\section{BPHZ scheme on the lattice} \label{sec:bphz}

Within the BPHZ scheme the ultraviolet divergences of integrals are removed
by performing subtractions directly in the integrand of a Feynman integral \cite{Zi}.
This is achieved by subtracting the first few terms of the Taylor expansion
in the external momenta. Originally this method was developed for the continuum
formulation of field theories, but afterwards extended to lattice field theories 
as well \cite{Re}.

\begin{figure}[t]
\centering\includegraphics[width=0.28\textwidth]{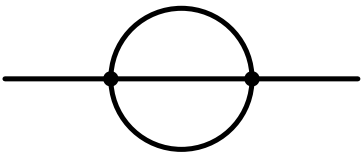}
\caption{Shown is the sunset diagram with three massive propagators.}
\label{fig:sunset}
\end{figure}

In the present work we are particularly interested in isolating the divergent pieces
of a given lattice loop integral, while the remaining finite pieces can be calculated
numerically for a given lattice spacing. 
We restrict ourselves to massive scalar field theories on the lattice. For illustration
the method will be worked out explicitly for the basic sunset diagram which arises
in $\phi^4$ theory at the two-loop level, see Fig.~\ref{fig:sunset}, 
but a general two-loop diagram can be evaluated in the same manner.
The first step in the approach is the application of the BPHZ scheme so that
the would-be divergences which occur as the lattice spacing goes to zero are contained
in lattice integrals without external momenta, while the momentum-dependent but finite remainder
can be neglected for renormalization purposes. 
Provided the lattice theory satisfies the power counting conditions given
in \cite{Re2}, which is the case here,
the continuum limit of these momentum-dependent finite pieces
exists, is independent of the details of the lattice action and is identical
with the BPHZ finite parts obtained from the corresponding continuum action \cite{Re}.
In order to illustrate the method, 
we will confirm explicitly for the case of the sunset diagram that this remainder is indeed finite
in the continuum limit. 
For definiteness we work with a hypercubical lattice $\Lambda$ of infinite volume
and the standard free scalar propagator,
while the results are presented in lattice units if not otherwise stated.

The basic sunset diagram for a massive scalar field with external momentum $p$ 
is given by Fig.~\ref{fig:sunset}, 
\beq \label{eq:sunset}
\int_{-\pi}^\pi  \frac{d^4 k}{(2 \pi)^4} \frac{d^4 q}{(2 \pi)^4} I_S(k,q;p)
= \int_{-\pi}^\pi  \frac{d^4 k}{(2 \pi)^4} \frac{d^4 q}{(2 \pi)^4} 
\frac{1}{\Delta(k)\Delta(q)\Delta(p+k+q)} \ ,
\eeq
where the subscript $S$ of $I$ denotes the full sunset diagram and $\Delta$ is the
inverse lattice propagator
\beq
\Delta(p) = m^2 + \hat{p}^2   =  m^2 + \sum_{\mu=1}^4 \hat{p}_\mu^2   =
 m^2 + 4 \sum_{\mu=1}^4 \sin^2 \left(\frac{ p_\mu}{2} \right)  \ .
\eeq

\begin{figure}[b]
\centering
\begin{minipage}[b]{0.22\textwidth}
  \includegraphics[width=\textwidth]{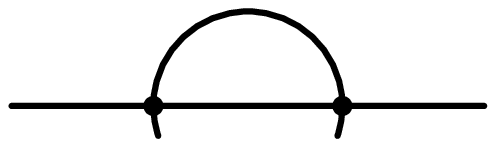}
  \vspace{0pt}
\end{minipage}
\hspace{0.5cm}
\begin{minipage}[b]{0.22\textwidth}
  \vspace{0pt}
  \includegraphics[width=\textwidth]{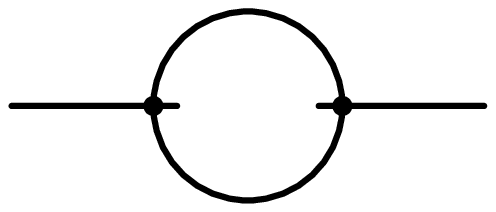} 
\end{minipage}
\hspace{0.5cm}
\begin{minipage}[t]{0.22\textwidth}
  \includegraphics[width=\textwidth]{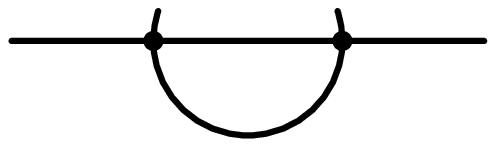}
  \vspace{0pt}
\end{minipage}
\caption{Subdiagrams $\gamma_{1,2,3}$ of the sunset diagram.}
\label{fig:subdia}
\end{figure}

Let $\gamma_1,\gamma_2,\gamma_3$ 
be the three subdiagrams of $S$, Fig.~\ref{fig:subdia}, then the corresponding set of forests is given
by ${\cal F}(S)= \{ \emptyset, S, \gamma_1,\gamma_2,\gamma_3,S\gamma_1,S\gamma_2,S\gamma_3 \}$
with the superficial degrees of divergence $d(S)=2$ and $d(\gamma)=0$ otherwise.
The integrand $I_S$ of the sunset diagram leads to would-be divergences in the continuum
limit as can be seen by explicitly inserting the lattice spacing in Eq.~(\ref{eq:sunset}).
A convenient renormalization procedure to handle these divergences is provided by
the forest formula of Zimmermann \cite{Zi} wherein a subtraction operation is directly applied  
to the integrand. Within this approach the renormalized Feynman integrand $R_S$ is given by
\beq
R_S = \sum_{\Gamma \in{\cal F}} \ \prod_{\gamma \in \Gamma} \ ( - t_\gamma^{d(\gamma)} ) I_S \, ,
\eeq
where $t_\gamma^{d(\gamma)}$ are lattice subtraction operators of order $d(\gamma)$ 
as introduced in \cite{Re}.
In the continuum limit the operators $t_\gamma^{d(\gamma)}$ coincide with
the first $d(\gamma)$ terms of the Taylor expansion
of the integrand $I_\gamma$ in the external momentum of the subgraph. For the sunset diagram one obtains
\beqa
R_S &=&  \Big( 1 - t_{S}^2 - t_{\gamma_1}^0 - t_{\gamma_2}^0 - t_{\gamma_3}^0 + t_S^2 \big[
          t_{\gamma_1}^0 + t_{\gamma_2}^0 + t_{\gamma_3}^0 \big]  \Big) I_S \no \\
    &=&  \Big( 1 - t_{S}^2 - [1 - t_S^2 ]\big[
          t_{\gamma_1}^0 + t_{\gamma_2}^0 + t_{\gamma_3}^0 \big]  \Big) I_S \ .
\eeqa
Since $[1 - t_S^2 ] \, ( t_{\gamma_1}^0 + t_{\gamma_2}^0 + t_{\gamma_3}^0 ) I_S =0$, this leads to
\beq  \label{eq:renorm}
R_S = [1 - t_S^2 ] \, I_S \ . 
\eeq
A convenient choice for the subtraction operator $t_S^2$ is given by \cite{Re} 
\beq
t_S^2 \, I_S(k,q;p) = I_S(k,q;0) + \sum_\mu \sin p_\mu \left( \frac{\partial}{\partial p_\mu} I_S \right)_{p=0}
    + \frac{1}{2} \sum_{\mu , \nu} \sin p_\mu \sin p_\nu \left( \frac{\partial}{\partial p_\mu} 
         \frac{\partial}{\partial p_\nu} I_S \right)_{p=0} \ ,
\eeq
where the partial derivatives are to be taken at $p=0$.
This yields
\beqa
t_S^2 \, I_S(k,q;p) &=& \Delta^{-1}(k)\Delta^{-1}(q)\Delta^{-1}(k+q) \Big( 1 -  \sum_\mu \sin^2 p_\mu
        \cos (k+q)_\mu  \Delta^{-1}(k+q) \no \\
       && \qq \qq \qq \qq \qq \q + 4  \sum_\mu \sin^2 p_\mu \, \sin^2 (k+q)_\mu  \Delta^{-2}(k+q)  \Big) \ .
\eeqa
Utilizing hypercubical symmetry of the lattice it can be rewritten as
\beqa
t_S^2 \, I_S(k,q;p) &=& \Delta^{-1}(k)\Delta^{-1}(q)\Delta^{-1}(k+q) \Big( 1 + \frac{1}{8} \bar{p}^2  
      - (1 +\frac{m^2}{8}) \bar{p}^2 \Delta^{-1}(k+q) \no \\
       && \qq \qq \qq \qq \qq \q + 4 \sum_\mu \bar{p}^2_\mu \, (\overline{k+q})^2_\mu \, \Delta^{-2}(k+q)  \Big) 
\eeqa
with 
\beq
\bar{p}^2 = \sum_{\mu =1}^4 \bar{p}^2_\mu = \sum_{\mu =1}^4 \sin^2 p_\mu \ .
\eeq
Neglecting pieces that do not lead to divergencies in the continuum limit
this simplifies to
\beqa  \label{eq:tgig}
t_S^2 \, I_S(k,q;p) &=& \Delta^{-1}(k)\Delta^{-1}(q)\Delta^{-1}(k+q) \Big( 1 
      - \bar{p}^2 \Delta^{-1}(k+q) \no \\
       && \qq \qq \qq \qq \qq \q + 4 \sum_\mu \bar{p}^2_\mu \, (\overline{k+q})^2_\mu \,  
       \Delta^{-2}(k+q)  \Big) + \ldots \quad  . 
\eeqa
According to \cite{Re} integration of $R_S$ in Eq.~(\ref{eq:renorm}) yields a
finite continuum limit which is independent of the details of the lattice action.
In order to demonstrate the finiteness of the renormalized integral, it is thus sufficient
to consider the corresponding integral in the continuum formulation 
\beqa \label{eq:contint}
&& \int_{-\infty}^\infty  \frac{d^4 k}{(2 \pi)^4} \frac{d^4 q}{(2 \pi)^4} R_S^{ct} (k,q;p) \no \\
  &&  
   = \int_{-\infty}^\infty  \frac{d^4 k}{(2 \pi)^4} \frac{d^4 q}{(2 \pi)^4} \Bigg\{
D^{-1}(k)D^{-1}(q)^{-1}D^{-1}(p+k+q) \no \\
  &&   - D^{-1}(k)D^{-1}(q)D^{-1}(k+q)   \Big( 1 
   -  p^2 D^{-1}(k+q) + 4 \sum_\mu p^2_\mu (k+q)^2_\mu  D^{-2}(k+q)  \Big)  \Bigg\}
\eeqa
with the inverse continuum propagator
\beq
D(p) = m^2 +  p^2 = m^2 + \sum_{\mu=1}^4 p_\mu^2 \ .
\eeq
We employ dimensional regularization for the evaluation of the Feynman integral (\ref{eq:contint}) 
which can then be simplified according to
\beqa \label{eq:dimreg}
&& \int_{-\infty}^\infty \frac{d^d k}{(2 \pi)^d} \frac{d^d q}{(2 \pi)^d} R_S^{ct} (k,q;p) \no \\
  &&  
   = \int_{-\infty}^\infty  \frac{d^d k}{(2 \pi)^d} \frac{d^d q}{(2 \pi)^d} \Bigg\{
D^{-1}(k)D^{-1}(q)^{-1}D^{-1}(p+k+q) \no \\
  &&   - D^{-1}(k)D^{-1}(q)D^{-1}(k+q)   \Big( 1 
   -  p^2 m^2 D^{-2}(k+q) +  \frac{\epsilon}{4} p^2 D^{-1}(k+q)  \Big)  \Bigg\}  + \ldots \quad ,
\eeqa
where $\epsilon = 4-d$ and we have neglected finite terms. One verifies in a straightforward manner
that the divergent pieces of the single terms in Eq.~(\ref{eq:dimreg}) indeed cancel.
This is most conveniently done by noting that in dimensional regularization the divergent pieces of
the sunset integral with two different masses,
\beq \label{eq:sunsetdiffmass}
I(m^2, m^2, M^2;p^2) = \int_{-\infty}^\infty  \frac{d^d k}{(2 \pi)^d} \frac{d^d q}{(2 \pi)^d} 
\frac{1}{[m^2+ k^2] \, [m^2+ q^2] \, [M^2+ (k+q+p)^2] } \ ,
\eeq
are given by 
\beq \label{eq:sunsetdiv}
I_{div}(m^2, m^2, M^2;p^2) =  \frac{1}{\epsilon -1} \ \frac{\Gamma(\epsilon)}{(4 \pi)^{4-\epsilon}} \
\Big[ 2 m^2 \, \big( \frac{2}{\epsilon} + 1 - 2 \ln m^2  \big) 
+ M^2 \, \big( \frac{2}{\epsilon} + 1 - 2 \ln M^2  \big) + \frac{p^2}{2}    \Big]   \ .
\eeq
The divergences of the scalar integrals in 
Eq.~(\ref{eq:dimreg}) can now be deduced directly from the expression in
Eq.~(\ref{eq:sunsetdiv}) and its differentiation with respect to $M^2$.
One obtains
\beq 
\int_{-\infty}^\infty \frac{d^d k}{(2 \pi)^d} \frac{d^d q}{(2 \pi)^d} 
R_S^{ct} (k,q;p) \Big\vert_{div}   = 0 \ .
\eeq
We have thus confirmed explicitly that the would-be divergences of the original
lattice integral (\ref{eq:sunset}) are contained in the expression $t_S^2 I_S$ of
Eq.~(\ref{eq:tgig}).
As there is no external momentum left in the propagators of $t_S^2 I_S$, 
one can restrict oneself in the following to loop integrals without
external momenta in the propagators.

\section{Coordinate space method} \label{sec:coord}

After having applied the BPHZ formalism to the original sunset integral we
now continue with the evaluation of the three lattice integrals appearing
in $\int t_S^2 I_S$, Eq.~(\ref{eq:tgig}). Instead of calculating the 
divergent pieces in momentum space, it is
more convenient to work in coordinate space. To this end, one can employ
the results for the massive Green's function presented in \cite{BK}.
We start with the integral
\beq  \label{eq:int1}
\int_{-\pi}^\pi  \frac{d^4 k}{(2 \pi)^4} \frac{d^4 q}{(2 \pi)^4} 
\frac{1}{\Delta(k)\Delta(q)\Delta(k+q)} = \sum_{x \in \Lambda} G(x)^3\  ,
\eeq
where $G$ is the lattice Green's function
\beq  
G(x)=\int_{-\pi}^\pi\frac{d^4k}{(2\pi)^4}
\frac{e^{i\,k\cdot x}}{\hat{k}^2+m^2}.
\eeq
In lattice units
the sum over $G^3$ is finite and can be carried out numerically
(with increasing precision for larger values of $m$)
utilizing the precise numerical knowledge of the Green's function \cite{BK}. 
In order to isolate the would-be divergences for $a \to 0$,
however, one recalls that in physical units the sunset diagram in Eq.~(\ref{eq:sunset}) 
enters with a prefactor
$1/a^2$, where $a$ is the lattice spacing and the physical mass $\mu$ is given by
$m = a \mu$. 
For large enough $x$ the first term in the small mass expansion of $G$, i.e. the massless
Green's function $G_{m=0}$, behaves asymptotically as $1/x^2$ and, hence, $G^3_{m=0}$ as $1/x^6$ \cite{LW}.
One can thus perform the sum explicitly. But for the logarithmic
divergences which arise in the subleading mass term one observes an
asymptotic $1/x^4$ behavior of the expansion coefficient of $G^3$ and the
summation over all lattice sites cannot be performed numerically. We therefore 
proceed by splitting the sum of the Green's functions as follows:
\beq  \label{eq:sumsplit}
\sum_{x \in \Lambda} G(x)^3 = G(0)^3 + \sum_{x \ne 0} (G(x)^3 - G_{as}^{(0)}(x)^3 )
+ \sum_{x \ne 0} G_{as}^{(0)}(x)^3 .
\eeq
In this expression, we have introduced the continuum Green's function
\beq
G_{as}^{(0)}(x) = \frac{1}{4\pi^2 x^2}m|x| K_1(m|x|) \ ,
\eeq
where $|x| = \sqrt{x^2}$ and $K_1$ represents the modified Bessel function of the second kind
\beq
K_\nu(z)=\frac{\pi}{2}\frac{I_{-\nu}(z)-I_\nu(z)}{\sin(\nu \pi)},\quad
I_\nu(z)=\sum_{i=0}^\infty\frac{(z/2)^{\nu+2i}}{i!\,\Gamma(\nu + i +1)} \ .
\eeq
The Green's function at the origin, $G(0)$, has been separated from the sum, as 
$G_{as}^{(0)}(x)$ diverges at the origin.
Note that we have merely subtracted (and added) the continuum
Green's function which is sufficient to expand the difference in the first sum on the right
side of Eq.~(\ref{eq:sumsplit}) up to order $m^2$. In order to be able to expand
to higher orders in $m^2$ one must subtract further asymptotic pieces of the lattice Green's
function which are suppressed by increasing powers in the lattice spacing
and represent corrections to the continuum expression. They have
already been presented in \cite{PS}, where the asymptotic behavior
of the lattice Green's function for large $x$ has been studied. For completeness,
we present in Appendix~\ref{app:asympt} 
an alternative and much simpler derivation of the asymptotic form of 
the lattice Green's function.

For large $x$ and $m=0$ the difference $G(x)^3 - G_{as}^{(0)}(x)^3$ vanishes as $1/x^8$
so that the small mass expansion can be performed up to $m^2$. 
To this end, we note that the small mass
expansion of the Green's function is given by \cite{BK}
\beq
G(x)=\sum_{i=0}^\infty a_i(x)m^{2i}+m^2\ln(m^2)\sum_{i=0}^\infty 
b_i(x)m^{2 i},
\eeq
which converges absolutely for $m<2$. The functions 
$a_i(x)$ can be calculated to high accuracy using a set of recursion
relations, while the $b_i(x)$ can be given in closed form 
\beq
b_0(x)=\frac{1}{16\pi^2},\quad b_1(x)=\frac{1}{128\pi^2}(x^2-1),\quad
b_2(x)=\frac{1}{3072\pi^2}((x^2)^2-4 x^2 + 3) , \quad  \mbox{etc.}
\eeq
Moreover, the asymptotic large-$x$ behavior of the coefficients $a_i$, $a_i^{as}$, can
be derived by expanding $G_{as}^{(0)}$ in $m$ 
\beqa \label{eq:coeffasy}
a_0^{as}(x)&=&\frac{1}{4\pi^2 x^2} + {\cal O}(1/x^4),\nonumber\\
a_1^{as}(x)&=&\frac{1}{16\pi^2}\Big[-1  
+ 2\Big(\gamma_E+\ln\Big(\frac{|x|}{2}\Big)\Big)\Big] + {\cal O}(1/x^2) 
\eeqa 
with $\gamma_E=0.5772\ldots$ being the Euler-Mascheroni constant.
(In Eq.~(\ref{eq:coeffasy}), subleading orders in $1/x^2$ can be deduced from the higher
order corrections of the asymptotic Green's function.)
We therefore obtain up-to-and-including order $m^2$ the expressions
\beq  \label{eq:g03}
G(0)^3 = a_0(0)^3 + 3 m^2 a_0(0)^2 a_1(0) + 3 m^2 \ln(m^2) a_0(0)^2 b_0(0) + {\cal O}(m^4)
\eeq
and
\beqa \label{eq:asym2}
\sum_{x \ne 0} (G(x)^3 - G_{as}^{(0)}(x)^3 ) &=&
\sum_{x \ne 0}  (a_0(x)^3 -  a_0^{as}(x)^3 ) + 3 \, m^2 \sum_{x \ne 0}  (a_0(x)^2 a_1(x) 
         -  a_0^{as}(x)^2 a_1^{as}(x)) \no \\
&& + 3 \, m^2 \ln (m^2) \sum_{x \ne 0}  (a_0(x)^2 b_0(x) 
         -  a_0^{as}(x)^2 b_0(x))  + {\cal O}(m^4) \no \\
         &=&
\sum_{x \ne 0}  \Big(a_0(x)^3 -  \frac{1}{64 \pi^6 x^6} \Big) \no \\
&&    + 3 \, m^2 \sum_{x \ne 0}      \bigg(  a_0(x)^2 a_1(x)      
       - \frac{1}{256 \pi^6 x^4 } \Big[ 2 \gamma_E+2\ln\Big(\frac{|x|}{2}\Big)-1 \Big] \bigg) \no \\
&& + 3 \, m^2 \ln (m^2) \sum_{x \ne 0}  \Big(a_0(x)^2 b_0(x) 
         -  \frac{1}{256 \pi^6 x^4 }\Big) + {\cal O}(m^4) \ .
\eeqa
In the continuum limit, the first and third sum in Eq.~(\ref{eq:asym2}) 
lead to quadratic and logarithmic 
divergences, respectively, whereas the second sum remains finite.

We can now turn our attention to the evaluation of the last term in Eq.~(\ref{eq:sumsplit})
\beq
\sum_{x \ne 0} G_{as}^{(0)}(x)^3 = \sum_{x \ne 0} \Big( \frac{1}{4\pi^2 x^2}m|x| K_1(m|x|) \Big)^3 
\eeq
which we rewrite as 
\beq  \label{eq:splitk1}
\sum_{x \ne 0} G_{as}^{(0)}(x)^3 =\sum_{x \ne 0} 
\bigg[\Big( \frac{1}{4\pi^2 x^2}m|x| 
K_{1}(m|x|) \Big)^3 - A(m,x,4) \bigg]
+ \sum_{x \ne 0}  A(m,x,4) \ .
\eeq
Here, $A(m,x,d)$ is given by
\beqa  
A(m,x,d) &=& \sum_{j=0}^3
\kappa_j\, C\left(3\left[\frac{d}{2}-1\right]-j,m\,x\right) + 
\sum_{j=0}^2\kappa_{j+4}\,C\left(2\left[\frac{d}{2}-1\right]-j,m\,x\right)
\nonumber\\
&+&
\sum_{j=0}^1\kappa_{j+7}\,C\left(\frac{d}{2}-1-j,m\,x\right),
\label{eq:Ad}
\eeqa
with 
\beq \label{eq:cfunc}
C(\nu, x) = \left(\frac{2}{x}\right)^\nu K_\nu(x),
\eeq
and the coefficients $\kappa_0,\dots,\kappa_8$, which also depend on $m$ and $d$, 
are given in Appendix~\ref{app:coeff}.
The function $A$ has been chosen in such a way that the difference
\beqa 
b&=&\lim_{x\to0} \bigg[\Big( \frac{1}{4\pi^2 x^2}
m|x| K_{1}(m|x|) \Big)^3 - A(m,x,4) \bigg]={\cal O}(m^4) \label{eq:difflim}
\eeqa
remains finite at the origin. We can thus replace the first sum in Eq.~(\ref{eq:splitk1}) 
by the sum over all lattice sites including $x=0$.
We are left with the evaluation of the difference on the right-hand side of 
Eq.~(\ref{eq:splitk1}) which is rewritten as 
\beq  \label{eq:diffd}
\sum_{x \in \Lambda} 
\bigg[\Big( \frac{1}{4\pi^2 x^2} m|x| 
K_{1}(m|x|) \Big)^3 - A(m,x,4) \bigg] \ - \ b \ .
\eeq
Taking in Eq.~(\ref{eq:diffd}) the Poisson resummation 
\beq
\sum_{x \in \Lambda} f(x) = \sum_{x \in \Lambda}  \int_{-\infty}^\infty  \frac{d^4 p}{(2 \pi)^4} 
e^{-i p \cdot x} \tilde{f} (p) = \sum_{k } \tilde{f} (2 \pi k)
\eeq
yields  
\beqa  \label{eq:gas03}
\sum_{x\neq0}G_{as}^{(0)}(x)^3 &=& \sum_{k}\lim_{d\to4}\Bigg[I(m^2,m^2,m^2;(2\pi k)^2)
-\tilde{A}(m,2 \pi k,d)\Bigg]-b\no\\
&& +
\sum_{x \ne 0}A(m,x,4)  + {\cal O}(m^4)  
\eeqa
with $I$ the continuum sunset integral in $d$-dimensional space defined in
Eq.~(\ref{eq:sunsetdiffmass}) and with external 
momentum $2 \pi k$, while $\tilde{A}$ is given by
\beqa  
\tilde{A}(m,p,d) &=& \frac{(4\pi)^{d/2}}{2}\Bigg[\sum_{j=0}^3
\kappa_j\,\Gamma(3+j-d)\frac{m^{6+2j-3d}}{(p^2+m^2)^{3+j-d}} + 
\sum_{j=0}^2\kappa_{j+4}\,\Gamma\left(2+j-\frac{d}{2}\right)\no\\
&&\times\frac{m^{4+2j-2d}}{(p^2+m^2)^{2+j-\frac{d}{2}}} +
\sum_{j=0}^1\kappa_{j+7}\,\Gamma(1+j)\frac{m^{2+2j-d}}{(p^2+m^2)^{1+j}}\Bigg] \ .
\label{eq:Adtilde}
\eeqa
We have analytically continued the expressions in the first sum of Eq.~(\ref{eq:gas03})
to $d$-dimensional space.
For $d=4$, the single terms in the difference of Eq.~(\ref{eq:splitk1}) diverge
at the origin and cannot be calculated separately, whereas for $d$ small enough
both terms remain finite and can be evaluated. Taking afterwards the $d \to 4$ limit
one confirms that the divergences in $d$ cancel out in the difference.
By continuing to
$d$ dimensions we have automatically introduced dimensionally regularized integrals. 
In the course of evaluating sunset diagrams with momenta $2 \pi k$ in dimensional regularization
it is more convenient to discuss the case with momentum $k =0$ separately. The results for external
momentum $k=0$ are presented, e.g., in \cite{Laporta,Pozzorini}
\beqa  
I(m^2,m^2,m^2;0)&=&-\frac{3\,m^2}{128\,\pi^4}\frac{1}{(d-4)^2}
+\frac{3\,m^2}{256\,\pi^4}\left(3-2\gamma_E-2\ln(m^2)+2\ln(4\pi)\right)
\frac{1}{d-4}\nonumber\\
&+&\frac{m^2}{256\pi^4}\bigg[-\frac{21}{2}-3 \gamma_E^2-\frac{\pi ^2}{4}-9 \ln(4 \pi )-3 \ln^2(4 \pi )+2 \sqrt{3} \text{Cl}_2\left(\frac{\pi }{3}\right)\nonumber\\
&+&\gamma_E (9+6 \ln(4 \pi ))+\ln(m^2) (9-6 \gamma_E+6 \ln(4 \pi ))-3 \ln^2(m^2)\bigg]\label{eq:sunset0}
\eeqa
with $\text{Cl}_2$ being the Clausen function.
For $k \ne 0$, on the other hand, one can expand in powers of $m^2/(2 \pi k)^2$ \cite{Laporta,Pozzorini}
\beqa  \label{eq:sunsetk}
I(m^2,m^2,m^2;p^2)&=&-\frac{3\,m^2}{128\,\pi^4}\frac{1}{(d-4)^2}+\frac{p^2-12 m^2 \ln(m^2)+6 m^2 (3-2 \gamma_E+2 \ln(4 \pi))}{512 \pi ^4}\frac{1}{d-4}
\nonumber\\
&+&\frac{p^2}{2048 \pi ^4}
 \Big[-13+4 \gamma_E-4 \ln(4 \pi )+
4 \ln(p^2)\Big]-\frac{m^2}{1024 \pi ^4}\Big[30+12\gamma_E (-3+\gamma_E) \nonumber\\
&+&\pi^2+36 \ln(4 \pi )+6 \big[\ln^2(m^2)-\ln^2(p^2)+
2 \ln(m^2) (-3+2 \gamma_E+\ln(p^2)\no\\
&-&2 \ln(4 \pi ))+2 \ln(4 \pi ) (-2 \gamma_E+\ln(4 \pi ))\big]\Big]+ m^2{\cal O}(m^2/p^2) \ .
\eeqa
For our purposes we only need to expand up to $m^2$. Note that in
the difference $I-\tilde{A}$ the $1/(d-4)$ divergences cancel and 
the difference $I-\tilde{A}$ scales like $1/k^6$ for large $k$ so that the 
sum over $k$ converges.

We are left with the evaluation of the sum $\sum\limits_{x\neq0}A(m,x,4)$
in Eq.~(\ref{eq:gas03}). To this aim, it is convenient to work with
the integral representation of the Bessel functions $K$ leading to, cf. Eq.~(\ref{eq:cfunc}),
\beq
C(\nu,m\,x)=\frac{1}{2\, m^{2\nu}}\int_0^\infty d\lambda \,\lambda^{-\nu-1}
\exp\left(-m^2\lambda-\frac{x^2}{4\lambda}\right).
\eeq
Since the function $A$ is a linear combination of the $C$ functions, it is sufficient
to demonstrate the calculation for the sum
\beq
\sum_{x\neq0}C(\nu,m\,x) \ ,
\eeq
or in more general---if we are interested in the calculation of $m^4$ and 
higher order corrections---terms of the form 
\beq
\sum_{x\neq0}h_D(x)C(\nu,m\,x) \ .
\eeq
Here, $h_D(x)$ is a harmonic homogeneous polynomial of degree D with the leading orders given by
\beq
h_0(x)=1,\quad h_4(x)=2 x^4-(x^2)^2,\quad h_6(x)=16 x^6-20 x^2 x^4+5(x^2)^3 , 
\quad \ldots \quad .
\eeq 
Following \cite{LW} we introduce the heat kernel
\beq
k(\lambda,h_D)=\sum_x h_D(x)e^{-\pi\,\lambda\,x^2}
\eeq
and rewrite the original sum in the form
\beq
\sum_{x\neq0}h_D(x)C(\nu,m\,x)=\frac{1}{2}\left(\frac{4\pi}{m^{2}}\right)^\nu
\int_0^\infty d\lambda\,\lambda^{-\nu-1}e^{-\frac{m^2}{4\pi}\lambda}
\left[k\left(1/\lambda,h_D\right)-h_D(0)\right].
\eeq
Poisson resummation and the fact that $h_D(x)$ is a harmonic homogeneous polynomial of 
degree $D$ lead to the relation
\beq
k(\lambda,h_D)=(-1)^{D/2}\lambda^{-D-2}k(1/\lambda,h_D).
\eeq
Employing this relation we can separate the regular and singular parts of the integral
\beqa
\sum_{x\neq0}h_D(x)C(\nu,m\,x)&=&\frac{1}{2}\left(\frac{4\pi}{m^{2}}\right)^\nu\int_1^\infty 
d\lambda\,\left[k\left(\lambda,h_D\right)-h_D(0)\right]\left[
(-1)^{D/2}e^{-\frac{m^2}{4\pi}\lambda}\lambda^{D-\nu+1}+e^{-\frac{m^2}{4\pi\lambda}}
\lambda^{\nu-1}\right]\no\\
&+&\frac{1}{2}\left(\frac{4\pi}{m^{2}}\right)^\nu h_D(0)\left[(-1)^{D/2}E_{\nu-D-1}\left(\frac{m^2}{4\pi}\right)-E_{\nu+1}\left(\frac{m^2}{4\pi}\right)\right],\label{sum:hC}
\eeqa
with the exponential integral functions $E_n$,
\beq
E_n(z)=\int_1^\infty d\lambda\, e^{-z\,\lambda}\lambda^{-n}.
\eeq
The first term in Eq.~(\ref{sum:hC}) is finite for all values of $\nu$ and 
can be Taylor expanded in $m^2$, while the exponential integral functions
entail the logarithmic terms.
The sum over the $A$ functions can now be 
represented as a linear combination of regularized zeta
functions and their derivatives with respect to $s$
\beq
\xi_R(s,h_D)=\int_1^\infty d\lambda\left[\lambda^{s-1}+(-1)^{D/2}\lambda^{D-s+1}\right]
\left[k(\lambda,h)-h_D(0)\right],\quad \xi_R^{(i)}(s,h_D)=\frac{d^i}{d s^i}\xi_R(s,h_D).
\eeq
The regularized zeta functions are related to the generalized zeta functions via
\beq
\xi(s,h_D) = \sum_{x \ne 0} h_D (x) (x^2)^{-s} = \frac{\pi^s}{\Gamma(s)}
\Big[ \frac{2 h_D(0)}{s (s-2)} + \xi_R(s,h_D) \Big] \ .
\eeq
Altogether, we obtain
\beqa
\sum_{x\neq0}A(m,x,4)&=&\frac{1}{(4\pi)^3}\left[\frac{1}{3}+\frac{\xi_R(3,h_0)}{2}\right]+m^2\frac{1}{(4\pi)^4}\left[\frac{11}{6}-\frac{3 \gamma_E}{2}-
\frac{3 \gamma_E^2}{2}+\frac{\pi ^2}{4}+\frac{3}{2} \ln(4 \pi )\right.\no\\
&+&\left.
3 \gamma_E \ln(4 \pi )-
\frac{3}{2} \ln^2(4 \pi )+3 \gamma_E \xi_R(2,h_0)-
3 \ln(4 \pi ) \xi_R(2,h_0)-3 \xi_R^{(1)}(2,h_0)\right]\no\\
&+&m^2\ln(m^2)\frac{1}{(4\pi)^4}\left[-\frac{3}{2}-3 \gamma_E+
3 \ln(4 \pi)+3 \xi_R(2,h_0)\right]-m^2\ln^2(m^2)\frac{3}{2}\frac{1}{(4\pi)^4}\no\\&+&{\cal O}(m^4).\label{eq:Amx4}
\eeqa

We can now summarize the quadratic, logarithmic and bilogarithmic divergences as well as
the finite part of the
integral in Eq.~(\ref{eq:int1}).
From Eqs.~(\ref{eq:g03}), (\ref{eq:asym2}),(\ref{eq:sunset0}),(\ref{eq:sunsetk})and (\ref{eq:Amx4}) we obtain the quadratic would-be divergences
\beqa  \label{eq:logdiv}
&&\int_{-\pi}^\pi  \frac{d^4 k}{(2 \pi)^4} \frac{d^4 q}{(2 \pi)^4} 
\frac{1}{\Delta(k)\Delta(q)\Delta(k+q)} \Big \vert_{\mbox{\scriptsize quadr.div.}} \no \\
&& =a_0(0)^3+\sum_{x \ne 0}  \Big(a_0(x)^3 -  \frac{1}{64 \pi^6 x^6} \Big)
+\frac{1}{(4\pi)^3}\left[\frac{1}{3}+\frac{\xi_R(3,h_0)}{2}\right]
\eeqa
the logarithmic divergences
\beqa
&&\int_{-\pi}^\pi  \frac{d^4 k}{(2 \pi)^4} \frac{d^4 q}{(2 \pi)^4} 
\frac{1}{\Delta(k)\Delta(q)\Delta(k+q)} \Big \vert_{\mbox{\scriptsize log.div.}}   \no \\
&& =m^2 \ln(m^2) \bigg[3 a_0(0)^2 b_0(0) +  \sum_{x \ne 0}  \Big(3 a_0(x)^2 b_0(x) 
         -  \frac{3}{256 \pi^6 x^4 }\Big) + \frac{1}{(4\pi)^4}\Big(-\frac{3}{2}-3 \gamma_E\no\\
&&+
3 \ln(4 \pi )+3 \xi_R(2,h_0)\Big)   \bigg] \ , \label{scalar:log}
\eeqa
the bilogarithmic divergences
\beq
\int_{-\pi}^\pi  \frac{d^4 k}{(2 \pi)^4} \frac{d^4 q}{(2 \pi)^4} 
\frac{1}{\Delta(k)\Delta(q)\Delta(k+q)} \Big \vert_{\mbox{\scriptsize bilog.div.}} 
= -\frac{3}{2(4\pi)^4} m^2 \, \ln^2(m^2) \ ,\label{scalar:bilog}
\eeq
as well as the finite remainder in the continuum limit
\beqa
&&\int_{-\pi}^\pi  \frac{d^4 k}{(2 \pi)^4} \frac{d^4 q}{(2 \pi)^4} 
\frac{1}{\Delta(k)\Delta(q)\Delta(k+q)} \Big \vert_{\mbox{\scriptsize finite}}   \no \\
&& =m^2\bigg[3a_0(0)^2a_1(0)+\sum_{x\ne0}\big(3a_0(x)^2a_1(x)-
\frac{3}{256\pi^6x^4}(2\gamma_E+2\ln(x/2)-1)\big)+\frac{1}{(4\pi)^4}\bigg(-\frac{15}{4}\no\\
&&-\frac{3 \gamma_E}{2}-
\frac{3 \gamma_E^2}{2}
+\frac{\pi ^2}{4}+2 \sqrt{3} \text{Cl}_2\left(\frac{\pi }{3}\right)+
\frac{3}{2} \ln(4 \pi )+3 \gamma_E \ln(4 \pi )-
\frac{3}{2} \ln^2(4 \pi )+3 \gamma_E \xi_R(2,h_0)\no\\
&&-
3 \ln(4 \pi ) \xi_R(2,h_0)
-3 \xi_R^{(1)}(2,h_0)\bigg)\bigg] \ . \label{scalar:finite}
\eeqa

So far, we have evaluated the divergent components of the sunset integral given
in Eq.~(\ref{eq:int1}). The calculation of the divergences in the second integrand
in $t_S^2 I_S$, Eq.~(\ref{eq:tgig}), is immediately obtained via
\beq \label{eq:int2}
\int_{-\pi}^\pi  \frac{d^4 k}{(2 \pi)^4} \frac{d^4 q}{(2 \pi)^4} 
\frac{1}{\Delta(k)\Delta(q)\Delta(k+q)^2} =
- \frac{1}{3} \frac{d}{d m^2} \int_{-\pi}^\pi  
\frac{d^4 k}{(2 \pi)^4} \frac{d^4 q}{(2 \pi)^4} 
\frac{1}{\Delta(k)\Delta(q)\Delta(k+q)}
\eeq
and by noting that this time we only need the small mass expansion at leading order due to the
presence of the prefactor $\bar{p}^2$ in Eq.~(\ref{eq:tgig}). Thus, the second term in
$t_S^2 I_S$ contributes only to the logarithmic divergences.

The third and final contribution in $t_S^2 I_S$, Eq.~(\ref{eq:tgig}), is given
by  
\beq    \label{eq:int3}
4 \int_{-\pi}^\pi  \frac{d^4 k}{(2 \pi)^4} \frac{d^4 q}{(2 \pi)^4} 
\frac{\sin^2(k_\mu + q_\mu)}{\Delta(k)\Delta(q)\Delta(k+q)^3} =
- \frac{1}{4}\sum_x G_1(x)^2 \sum_\mu[\nabla_\mu+ \nabla_\mu^*]^2 G_3(x) \ .
\eeq
Here, we have introduced the notation
\beq
G_i(x) = \int_{-\pi}^\pi  \frac{d^4 k}{(2 \pi)^4}  \frac{e^{ik \cdot x}}{[\Delta(k)]^i }
\eeq
and the backward and forward lattice derivatives are defined by
\beq
\nabla_\mu^* f(x)= f(x) - f(x-\mu)  \ , \qq  \nabla_\mu f(x)= f(x+\mu) - f(x) \ .
\eeq
One easily verifies that
the most general lattice integral with one mass arising in the BPHZ subtraction procedure has the form
\beq    \label{eq:bphzgen}
\int_{-\pi}^\pi  \frac{d^4 k}{(2 \pi)^4} \frac{d^4 q}{(2 \pi)^4} 
\frac{\prod_\mu \hat{k}_\mu^{2l_\mu^1}   \,  \hat{q}_\mu^{2l_\mu^2}
  \widehat{(k+q)}_\mu^{2l_\mu^3}}{
\Delta(k)^\alpha\Delta(q)^\beta \Delta(k+q)^\gamma}  \ .
\eeq
From Eq.~(\ref{eq:int3}) it is obvious how a generic integral of the form (\ref{eq:bphzgen}) can
be treated with the proposed coordinate space method. The presented approach is
therefore suited to evaluate  {\it any} two-loop integral without external momentum
which arises in the BPHZ procedure. The extension of this method to scalar field
theories with different masses is also straightforward.

We proceed as above by separating in Eq.~(\ref{eq:int3}) 
the contribution at the origin from the remaining sum
and isolating the corresponding continuum result 
\beqa   \label{eq:diffint3}
&&-\frac{1}{4}  \sum_x G_1(x)^2 \sum_\mu[\nabla_\mu+ \nabla_\mu^*]^2 G_3(x) =
-\frac{1}{4} G_1(0)^2 \sum_\mu[\nabla_\mu+ \nabla_\mu^*]^2 G_3(0) - \sum_{x \ne 0} \Big[\frac{1}{4}  
   G_1(x)^2\no\\
&&\times \sum_\mu[\nabla_\mu+ \nabla_\mu^*]^2 G_3(x) 
- {G_1^c(x)}^2\sum_\mu \left(\frac{\partial}{\partial x_\mu}\right)^2 G_3^c(x)  \Big]
 - \sum_{x \ne 0} {G_1^c(x)}^2 \sum_\mu\left(\frac{\partial}{\partial x_\mu}\right)^2 G_3^c(x)
\eeqa
with
\beq  \label{eq:greenspow}
G_i^c(x) = \int_{-\infty}^\infty  \frac{d^4 k}{(2 \pi)^4}  \frac{e^{i k \cdot x}}{[D(k)]^i } \ .
\eeq
(Note that in our notation $G_1^c = G_{as}^{(0)}$.)
The first term on the right side of Eq.~(\ref{eq:diffint3}) can be directly derived
from the Green's function at the origin \cite{BK},
and the subsequent term can also be calculated numerically, since the difference of functions in
the sum behaves like $1/x^6$ for large $x$ in the $m \to 0$ limit. 
In contrast, the rightmost term in Eq.~(\ref{eq:diffint3}) must be further decomposed
\beqa  \label{eq:int3split}
&&\sum_{x \ne 0} {G_1^c(x)}^2\sum_\mu \left(\frac{\partial}{\partial x_\mu}\right)^2 G_3^c(x) 
\no \\
&& =
\sum_{x} \Big[ {G_1^c(x)}^2 \sum_\mu\left(\frac{\partial}{\partial x_\mu}\right)^2 
              G_3^c(x) - B(m,x,4) \Big] \no \\
&& -  \lim_{x \to 0} \Big(  {G_1^c(x)}^2 
                 \sum_\mu \left(\frac{\partial}{\partial x_\mu}\right)^2 G_3^c(x) 
               - B(m,x,4)   \Big) + \sum_{x \ne 0} B(m,x,4) \no \\
&& =
\sum_{k} \int d^4 x \,  e^{i 2 \pi  k \cdot x}
  \Big[ {G_1^c(x)}^2 \sum_\mu\left(\frac{\partial}{\partial x_\mu}\right)^2 G_3^c(x) - B(m,x,4) \Big] \no \\
&& - \lim_{x \to 0} \Big(  {G_1^c(x)}^2 
 \sum_\mu \left(\frac{\partial}{\partial x_\mu}\right)^2 G_3^c(x) 
               - B(m,x,4)   \Big) + \sum_{x \ne 0} B(m,x,4) \ , 
\eeqa               
where the function $B(m,x,d)$ is defined as
\beqa  
B(m,x,d) &=& \sum_{j=0}^3
\lambda_j\, C\left(3\left[\frac{d}{2}-1\right]-j,m\,x\right) + 
\sum_{j=0}^2\lambda_{j+4}\,C\left(2\left[\frac{d}{2}-1\right]-j,m\,x\right)
\nonumber\\
&+&
\sum_{j=0}^1\lambda_{j+7}\,C\left(\frac{d}{2}-1-j,m\,x\right)
\label{eq:Bd}
\eeqa
with coefficients $\lambda_j$, which also depend on $m$ and $d$, 
given in Appendix~\ref{app:coeff}. The function $B(m,x,d)$
has been introduced in order to apply the Poisson summation formula.
The difference at the origin $x=0$ in Eq.~(\ref{eq:int3split}) is given by

\beqa
\lim_{x \to 0} \Big(  {G_1^c(x)}^2 \sum_\mu\left(\frac{\partial}{\partial 
              x_\mu}\right)^2 G_3^c(x) 
               - B(m,x,4)   \Big) &=& {\cal O}(m^4).   \no 
\eeqa
For the evaluation of the first term in Eq.~(\ref{eq:int3split}), on the other hand, we utilize
\beqa
&& \int d^d x \,  e^{i p \cdot x} {G_1^c(x)}^2 \sum_\mu\left(\frac{\partial}{\partial 
             x_\mu}\right)^2 G_3^c(x)
= \int d^d x \,  e^{i p \cdot  x} {G_1^c(x)}^2  \left( \frac{d}{4} \frac{\partial}{\partial m^2} 
    + \frac{x^2}{8} \right) G_1^c(x)\no\\
&&=\left[\frac{d}{12} \frac{\partial}{\partial m^2}- \frac{1}{8}  
  \left( \frac{\partial^2}{\partial |p|^2} 
    + \frac{d-1}{|p|} \frac{\partial}{\partial |p|}  \right)  
     \right] \int d^d x \,  e^{i p \cdot  x} {G_1^c(x)}^3 \ , 
\eeqa
where we introduced analytical continuation to $d$ dimensions in order to 
evaluate both terms in the first sum of Eq.~(\ref{eq:int3split}) separately.
For $k \ne 0$ we can directly apply the differential
operators of the last equation to the result of Eq.~(\ref{eq:sunsetk}). For $k=0$,
on the other hand, we apply
the differential operators to the small $p$ expansion of the continuum sunset
diagram. The leading order of this expansion has already been presented in
Eq.~(\ref{eq:sunset0}), while the $p^2$ terms are given by
\beq
p^2 \Big[\frac{1}{512 (-4+d) \pi ^4}+\frac{-27+36 \gamma_E+36 \ln(m^2)-36 \ln(4 \pi )+32 \sqrt{3}\, \text{Cl}_2\left(\frac{\pi
}{3}\right)}{18432 \pi ^4}\Big].
\eeq
Note that for our considerations we can neglect $p^4$ and higher contributions.
The Poisson resummation for $B$ follows immediately by noting that $B$ has the
same structure as $A$ in Eq.~(\ref{eq:Ad}) and replacing $\kappa_i$ by $\lambda_i$ in
Eq.~(\ref{eq:Adtilde}).
As final result we obtain 
\beqa
&& \int d^4 x \, \Big[{G_1^c(x)}^2 \sum_\mu\left(\frac{\partial}{\partial x_\mu}\right)^2 
G_3^c(x)-B(m,x,4)\Big] 
= \frac{1}{(4\pi)^4}\Big[-\frac{9}{4}+\frac{2}{3\sqrt{3}}\text{Cl}_2\left(\frac{\pi}{3}\right)\Big].
\eeqa

Summarizing the divergent and finite portions of the integral in 
Eq.~(\ref{eq:int3}) which remain in the continuum limit one obtains 
\beqa
&&4 \int_{-\pi}^\pi  \frac{d^4 k}{(2 \pi)^4} \frac{d^4 q}{(2 \pi)^4} 
\frac{\sin^2(k_\mu + q_\mu)}{\Delta(k)\Delta(q)\Delta(k+q)^3} 
\bigg \vert_{\mbox{\scriptsize log.div.}}  \no \\
&&= \ln(m^2)\bigg\{-\frac{1}{16\pi^2}a_0(0)^2+\sum_{x\neq0}\bigg[- \frac{1}{4}a_0(x)^2\sum_\mu 
\left[\nabla_\mu+\nabla_\mu^*\right]^2 b_1(x)
+\frac{1}{256\pi^6 x^4}\bigg]\no\\
&&+\frac{1}{(4\pi)^4}\big(2+\gamma_E-\ln(4 \pi )-\xi_R(2,h_0)\big)\bigg\},\label{tensor:log}\\[0.3cm]
&&4 \int_{-\pi}^\pi  \frac{d^4 k}{(2 \pi)^4} \frac{d^4 q}{(2 \pi)^4} 
\frac{\sin^2(k_\mu + q_\mu)}{\Delta(k)\Delta(q)\Delta(k+q)^3} \bigg 
\vert_{\mbox{\scriptsize bilog.div.}}  
= \frac{1}{2(4\pi)^4}\ln^2(m^2) \ , \label{tensor:bilog}
\eeqa
\beqa
&&4 \int_{-\pi}^\pi  \frac{d^4 k}{(2 \pi)^4} \frac{d^4 q}{(2 \pi)^4} 
\frac{\sin^2(k_\mu + q_\mu)}{\Delta(k)\Delta(q)\Delta(k+q)^3} \bigg 
\vert_{\mbox{\scriptsize finite}}  
=-a_0(0)^2\Big[\frac{3}{32 \pi ^2}+2 (a_2(2,0,0,0)-a_2(0))\Big]\no\\
&&+\sum_{x\neq0}\bigg[\frac{1}{512\pi^6 x^4}\big(1+4\gamma_E+4\ln(x/2)\big)
-\frac{1}{4} a_0(x)^2\sum_\mu \left[\nabla_\mu+\nabla_\mu^*\right]^2\left( a_2(x)
+\frac{3}{2}b_1(x)\right)\bigg]\no\\
&&+\frac{1}{(4\pi)^4}\Big[\frac{5}{2}+2 \gamma_E+\frac{\gamma_E^2}{2}-
\frac{\pi ^2}{12}-\frac{2 \text{Cl}_2\left(\frac{\pi }{3}\right)}{3 \sqrt{3}}-2 \ln(4 \pi )-
\gamma_E \ln(4 \pi )+\frac{1}{2} \ln^2(4 \pi )-
\frac{3 \xi_R(2,h_0)}{2}\no\\
&&-\gamma_E \xi_R(2,h_0)+
\ln(4 \pi ) \xi_R(2,h_0)+\xi_R^{(1)}(2,h_0)\Big] . \label{tensor:finite}
\eeqa
%

\section{Numerical results} \label{sec:numerics}

The Equations
(\ref{eq:logdiv}-\ref{scalar:finite}) and (\ref{tensor:log}-\ref{tensor:finite}) 
can in principle be evaluated numerically.
However, the respective sums converge slowly.
In order to increase the precision 
of the numerical results one must therefore accelerate the convergence of the sums. 
This can be achieved, e.g., by using the zeta function technique
introduced in \cite{LW}.

The small mass expansions of the integrals in Eqs.~(\ref{eq:sunset}) and (\ref{eq:int3}) are given by
\beqa
&&\int_{-\pi}^\pi  \frac{d^4 k}{(2 \pi)^4} \frac{d^4 q}{(2 \pi)^4} 
\frac{1}{\Delta(k)\Delta(q)\Delta(k+q)}=\sum_{i=0}^\infty c_i m^{2i}+m^2\ln(m^2)
\sum_{i=0}^\infty d_i m^{2i}+m^2\ln^2(m^2)\sum_{i=0}^\infty e_i m^{2i}  ,\no\\
&&4 \int_{-\pi}^\pi  \frac{d^4 k}{(2 \pi)^4} \frac{d^4 q}{(2 \pi)^4} 
\frac{\sin^2(k_\mu + q_\mu)}{\Delta(k)\Delta(q)\Delta(k+q)^3}=\sum_{i=0}^\infty 
f_i m^{2i}+\ln(m^2)\sum_{i=0}^\infty g_i m^{2i}+\ln^2(m^2)\sum_{i=0}^\infty h_i m^{2i} \ . \no
\eeqa
Recall that the expansion for the integral in Eq.~(\ref{eq:int2}) follows directly from the first
equation above by differentiation with respect to $m^2$.
Utilizing the zeta function method we have calculated the first few coefficients of these series.
We obtain
\beqa
c_0&=&0.0040430548122\ldots, \quad c_1=-0.0024114634124\ldots, \no \\
d_0&=&0.0006968046967\ldots,
\quad e_0=-\frac{3}{2}\frac{1}{(4\pi)^4} \ , \no
\eeqa
and
\beq \label{eq:numval}
f_0=0.0000731523655\ldots, \quad g_0=-0.0001721159925\ldots,\quad h_0=\frac{1}{2}\frac{1}{(4\pi)^4}.
\eeq
In principle, we can calculate higher coefficients with this method, but
for renormalization purposes this is not necessary. Also, the precision of the coefficients
can easily be improved by including higher orders in the zeta function technique and 
increasing computing time a bit.

\section{Additional relations for logarithmic terms}   \label{sec:relations}

The logarithmic terms can, in fact, be related to one-loop tadpoles.
Starting from the identity
\beq
\sum_x e^{i\, p\, x}G(x)^2 =\int_{-\pi}^\pi\frac{d^4k}{(2\pi)^4}
\frac{1}{\Delta(k)\Delta(p+k)}. 
\eeq
one has for $p=0$
\beq
\sum_x G(x)^2 =-\frac{\partial}{\partial m^2}G(0).
\eeq
On the other hand, decomposition of the sum
leads to
\beq
\sum_x G(x)^2=a_0(0)^2 + \sum_{x\neq0}\Big(a_0(x)^2-\frac{1}{16\pi^4 x^4}\Big)
+\sum_{x\neq0}G_{as}(x)^2+{\cal O}(m^2).
\eeq
Employing the small mass expansion of the tadpole
\beq
\frac{\partial}{\partial m^2}G(0)=a_1(0)+b_0(0)+b_0(0)\ln(m^2)+{\cal O}(m^2)
\eeq
one gets
\beq
\sum_{x\neq0}\Big(a_0(x)^2-\frac{1}{16\pi^4 x^4}\Big)=-a_0(0)^2
-\sum_{x\neq0}G_{as}(x)^2-a_1(0)-b_0(0)-b_0(0)\ln(m^2)+{\cal O}(m^2) .
\eeq
The sum over the asymptotic Green's function on the right-hand side can be calculated 
with the method presented in Sec.~\ref{sec:coord}:
\beq
\sum_{x\neq0}G_{as}(x)^2=\frac{1}{(4\pi)^2}\Big[-\frac{5}{2}-\gamma_E-
\ln(m^2)+\ln(4 \pi)+\xi_R(2,h_0)\Big]+{\cal O}(m^2).
\eeq
Comparison with Eq.~(\ref{scalar:log}) leads immediately to the logarithmic divergence
\beqa
&&\int_{-\pi}^\pi  \frac{d^4 k}{(2 \pi)^4} \frac{d^4 q}{(2 \pi)^4} 
\frac{1}{\Delta(k)\Delta(q)\Delta(k+q)} \Big \vert_{\mbox{\scriptsize log.div.}} 
=m^2 \ln(m^2)\frac{3}{(4\pi)^4} [1-(4\pi)^2 a_1(0)]
\eeqa
which reproduces the value for $d_0$ given in Eq.~(\ref{eq:numval})
when using $a_1=-0.030345...$ \cite{BK}.
By similar considerations we obtain for the logarithmic term of the tensor integral
\beqa
&&4 \int_{-\pi}^\pi  \frac{d^4 k}{(2 \pi)^4} \frac{d^4 q}{(2 \pi)^4} 
\frac{\sin^2(k_\mu + q_\mu)}{\Delta(k)\Delta(q)\Delta(k+q)^3} 
\bigg \vert_{\mbox{\scriptsize log.div.}}=\ln(m^2)\frac{1}{(4\pi)^4} 
\bigg[\frac{1}{2}+(4\pi)^2 a_1(0)\bigg]
\eeqa
which again is in agreement with $g_0$ from Eq.~(\ref{eq:numval}).
The additional relations derived here are not restricted to the 
leading order in the small mass expansion and one can obtain similar relations 
for higher order logarithmic pieces. 
In the remainder of this section, we will briefly illustrate the general strategy in
relating higher order logarithmic terms of two-loop diagrams to one-loop tadpole integrals.
For brevity, we will restrict ourselves to $\sum_x G(x)^3$, but the generalization
to tensor integrals is straightforward. We decompose $\sum_x G(x)^3$ as
\beq
\sum_x G(x)^3=G(0)^3+\sum_{x\neq0}\bigg[G(x)^3-F(x)\bigg]+\sum_{x\neq0}F(x),
\eeq
where the function $F(x)$ is chosen in such a way that the difference in the 
sum can be trivially expanded in the small mass up to a given order. The small mass expansion
of $F(x)$ reads
\beq
F(x)=\sum_{j=0}^3\sum_{i=0}^\infty f_i^{(j)}(x)m^{2i}[m^2\ln(m^2)]^j
\eeq 
with analytically known expansion coefficients $f_i^{(j)}$.
For the $n$-th order logarithmic term of the difference we get
\beq
\sum_{x\neq0}\bigg[G(x)^3-F(x)\bigg]_{m^{2n}\ln(m^2)}=
m^{2n}\ln(m^2)\sum_{x\neq0}\bigg[\sum_{i+j+k=n-1}b_i(x)a_j(x)a_k(x)-f_{n-1}^{(1)}(x) \bigg].\label{eq:nthLog}
\eeq
On the other hand, one has
\beq
\sum_x x^{2i} G(x)^2=\delta_{i,0}G(0)^2+\sum_{x\neq0}\bigg[x^{2i}G(x)^2-H(x)\bigg]+\sum_{x\neq0}H(x),
\eeq
where again the function $H(x)$ is chosen in such a way that the difference in the sum can be 
expanded in the small mass up to a given order. If the small mass
expansion of $H(x)$ is given by
\beq
H(x)=\sum_{j=0}^2\sum_{i=0}^\infty h_i^{(j)}(x)m^{2i}[m^2\ln(m^2)]^j \ ,
\eeq
then the $l$-th order of the difference reads
\beq
\sum_{x\neq0}\bigg[x^{2i}G(x)^2-H(x)\bigg]_{m^{2 l}}=m^{2 l}\sum_{x\neq0}\bigg[
\sum_{j+k=l}x^{2i}a_j(x)a_k(x)-h_{l}^{(0)}(x)\bigg].
\eeq 
Since $b_i(x)$ is a polynomial in $x$ we can always describe the lattice sum
of the logarithmic term in Eq.~(\ref{eq:nthLog}) as a linear combination of the terms
\beq
\sum_{x\neq0}\bigg[x^{2i}G(x)^2-H(x)\bigg]_{m^{2 l}}
\eeq 
and get in this way relations 
between the logarithmic pieces of $\sum_x G(x)^3$ and terms of the
type $\sum_x x^{2i_1}\dots x^{2i_n} G(x)^2$.
The latter terms can in turn be written as linear combinations of tadpoles with tensor
structures
\beqa
&&\sum_x x^{2i_1}\dots x^{2i_n} G(x)^2\no\\
&&=\Big[(-1)^{i_1}\sum_\mu\Big(\frac{\partial}{\partial p_\mu}\Big)^{2i_1}\Big]
\dots\Big[(-1)^{i_n}\sum_\mu\Big(\frac{\partial}{\partial p_\mu}\Big)^{2i_n}\Big] \Big\vert_{p=0}
\int_{-\pi}^\pi\frac{d^4k}{(2\pi)^4}\frac{1}{\Delta(k)}\frac{1}{\Delta(p+k)} 
\eeqa
such that the logarithmic pieces in $\sum_x G(x)^3$ are expressed in terms
of (tensor-)tadpole integrals.

\section{Conclusions} \label{sec:concl}

In this work, we have extracted the logarithmic and quadratic divergences of
the basic sunset diagram for a massive scalar field on the lattice and calculated
the coefficients of its would-be divergences to very high precision.
As a first step, the BPHZ fomalism has been applied to the sunset integral
leading to three integrals without external momentum in the propagators.
The quadratic and (bi-) logarithmic divergences are then extracted by applying coordinate space
techniques. 

The crucial observation is that the integrals can be written as products of
Green's functions and a summation over lattice sites is performed.
By subtracting the leading terms of the asymptotic form of the Green's function
one can expand the products of Green's functions in the mass $m$ of the scalar field
up to any given order. 
The coefficients of the logarithmic and quadratic divergences are thus
expressed in terms of sums over lattice sites.
Precise numerical knowledge of the Green's function values close to the origin
and of its asymptotic large-$x$ behavior make an accurate evaluation of these
sums possible.
Once the external momenta have been eliminated from the propagators (e.g. by applying
the BPHZ scheme), the method proposed can be utilized to calculate {\it any} two-loop 
diagram to very high accuracy.

\section*{Acknowledgments}
We thank Randy Lewis for useful discussions. 
We also thank the referee of the first manuscript version for a remark 
which triggered off the investigation in Sec.~\ref{sec:relations}.
Financial support of the
Deutsche Forschungsgemeinschaft is gratefully acknowledged.




\begin{appendix}
\section{Asymptotic expansion of $\mbox{\boldmath$G(x)$}$ for large $\mbox{\boldmath$x$}$
\label{app:asympt}}

In this appendix, we derive the asymptotic expansion of the lattice Green's function $G(x)$
for large~$x$. To this aim, it is convenient to introduce the lattice
constant $a$ and to study the expansion in $a$ with $\tilde{x}=a x$ fixed which
is equivalent to the large $x = \tilde{x}/a$ expansion. 

In \cite{PS}, Paladini and Sexton~ have studied the asymptotic
behavior of the quantity\footnote{In \cite{PS} the results were derived in $d$ dimensions, but
here we restrict ourselves to $d=4$ dimensions for simplicity.}
\beq
F_\alpha(x)=a^{4-2\alpha}\int_{-\pi/a}^{\pi/a}\frac{d^4k}{(2\pi)^4}e^{i\,k\cdot\tilde{x}}
\frac{1}{(\hat{k}^2+\mu^2)^\alpha},
\eeq
where $m=a\,\mu$, and in this appendix we set
\beq
\hat{k}^n=\sum_{\nu=1}^4 \left( \frac{2}{a}\sin \left(\frac{k_\nu a}{2} \right)  \right)^n.
\eeq

In the following, we present an alternative and much easier derivation of
their results. 
The small $a$ expansion of the integrand is given by
\beqa
\frac{1}{(\hat{k}^2+\mu^2)^\alpha}&=&\frac{1}{(k^2+\mu^2)^\alpha}
+a^2\frac{\alpha}{12}
\frac{k^4}{(k^2+\mu^2)^{\alpha+1}}
+a^4\left(\frac{\alpha(\alpha+1)}{288}\frac{(k^4)^2}{(k^2+\mu^2)^{\alpha+2}}
-\frac{\alpha}{360}\frac{k^6}{(k^2+\mu^2)^{\alpha+1}}\right)\nonumber\\
&+&a^6\left(\frac{\alpha(\alpha+1)(\alpha+2)}{10368}
\frac{(k^4)^3}{(k^2+\mu^2)^{\alpha+3}}-\frac{\alpha(\alpha+1)}{4320}
\frac{k^4 k^6}{(k^2+\mu^2)^{\alpha+2}}+\frac{\alpha}{20160}
\frac{k^8}{(k^2+\mu^2)^{\alpha+1}}\right)\nonumber\\
&+&{\cal O}(a^{8}), 
\eeqa 
such that the original integral reads
\beq
F_\alpha(x)\simeq F_\alpha^{(0)}(x)+a^2  F_\alpha^{(2)}(x)
+a^4  F_\alpha^{(4)}(x)+a^6  F_\alpha^{(6)}(x)+{\cal O}(a^{12- 2 \alpha})
\eeq
with
\beq
F_\alpha^{(0)}(x)=a^{4-2\alpha}\int_{-\pi/a}^{\pi/a}\frac{d^4k}{(2\pi)^4}e^{i\,k\cdot\tilde{x}}
\frac{1}{(k^2+\mu^2)^\alpha}\label{falpha0}
\eeq
and 
\beqa 
F_\alpha^{(2)}(x)&=&\frac{\alpha}{12}\partial_{\tilde{x}}^4 
F_{\alpha+1}^{(0)}(x),\nonumber\\
F_\alpha^{(4)}(x)&=&\frac{\alpha(\alpha+1)}{288}(\partial_{\tilde{x}}^4)^2
F_{\alpha+2}^{(0)}(x)+\frac{\alpha}{360}\partial_{\tilde{x}}^6
F_{\alpha+1}^{(0)}(x),\label{FAlpha2}\\
F_\alpha^{(6)}(x)&=&\frac{\alpha(\alpha+1)(\alpha+2)}{10368}
(\partial_{\tilde{x}}^4)^3 F_{\alpha+3}^{(0)}(x)+\frac{\alpha(\alpha+1)}{4320}
\partial_{\tilde{x}}^4\partial_{\tilde{x}}^6 F_{\alpha+2}^{(0)}(x)+
\frac{\alpha}{20160}\partial_{\tilde{x}}^8 F_{\alpha+1}^{(0)}(x).\nonumber
\eeqa
The derivative operators in Eqs.~(\ref{FAlpha2}) are defined by 
$\partial_{\tilde{x}}^n=\sum_{\nu=1}^4\partial^n/\partial\tilde{x}_\nu^n$.
The crucial observation is that
the extension of the integral boundaries in Eq.~(\ref{falpha0}) to infinity does not affect
the large-$x$ behavior of the integral but merely changes its behavior
near the origin. For our purposes we can thus set for $x \to \infty$
\beq
F_\alpha^{(0)}(x)=a^{4-2\alpha}\int_{-\infty}^{\infty}\frac{d^4k}{(2\pi)^4}e^{i\,k\cdot\tilde{x}}
\frac{1}{(k^2+\mu^2)^\alpha}=\left(\frac{a^2\mu}{|\tilde{x}|}\right)^{2-\alpha}
\frac{1}{2^{\alpha+1}\pi^2\Gamma(\alpha)}K_{2-\alpha}(|\tilde{x}|\,\mu),
\label{FAlpha1}
\eeq
where $|\tilde{x}|=\sqrt{\tilde{x}^2}$.  After
evaluating the derivatives in Eq.~(\ref{FAlpha2}) we arrive at the results of 
\cite{PS}. The asymptotic Green's function $G_{as}(x)$ is immediately obtained from
(\ref{FAlpha2}) by setting $\alpha=1$:
\beq
G_{as}(x)\simeq G_{as}^{(0)}(x)+G_{as}^{(2)}(x)+G_{as}^{(4)}(x)+G_{as}^{(6)}(x)+{\cal O}(a^{10})
\eeq
with
\beqa
G_{as}^{(0)}(x)&=&\frac{1}{4\pi^2 x^2}m|x| K_1(m|x|),\nonumber\\
G_{as}^{(2)}(x)&=&\frac{1}{96\pi^2 (x^2)^2}\Big[12 ( m |x|)^2 K_2(m |x|)
-6 (m |x|)^3 K_3(m |x|)+(m |x|)^4 \frac{x^4}{(x^2)^2}K_4(m |x|)\Big],
\nonumber\\
G_{as}^{(4)}(x)&=&\frac{1}{23040\pi^2 (x^2)^3}\Big[2160 ( m |x|)^3 K_3(m |x|)
-2280 ( m |x|)^4 K_4(m |x|) \nonumber\\
&+&\left(180  + 840 \frac{x^4}{(x^2)^2}\right)( m |x|)^5 K_5(m |x|)
-\left(60\frac{x^4}{(x^2)^2} + 72\frac{x^6}{(x^2)^3}\right)
(m |x|)^6 K_6(m |x|)\nonumber\\
&+&5 (m |x|)^7 \frac{(x^4)^2}{(x^2)^4}K_7(m |x|)\Big] \ ,\nonumber \\
G_{as}^{(6)}(x)&=&\frac{1}{11612160\pi^2(x^2)^4}\Big[
1179360 (m |x|)^4 K_4(m |x|)
-1935360 (m |x|)^5 K_5(m |x|)\nonumber\\ 
&+&\Big(241920 + 1270080 \frac{x^4}{(x^2)^2}\Big)(m |x|)^6 K_6(m |x|)
-\Big(7560 + 138600 \frac{x^4}{(x^2)^2}\nonumber\\
&+&\left.260064 \frac{x^6}{(x^2)^3}\Big)(m |x|)^7 K_7(m |x|)
+\Big(3780 \frac{x^4}{(x^2)^2} + 16380 \frac{(x^4)^2}{(x^2)^4} 
+ 9072 \frac{x^6}{(x^2)^3}\right.\nonumber\\
&+&\left. 16200 \frac{x^8}{(x^2)^4}\Big)(m |x|)^8 K_8(m |x|)
-\Big(630 \frac{(x^4)^2}{(x^2)^4}+1512 \frac{x^4 x^6}{(x^2)^5}\Big)
(m |x|)^9K_9(m |x|)\right.\nonumber\\
&+&\left.35 \frac{(x^4)^3}{(x^2)^6} (m |x|)^{10} K_{10}(m |x|)\right) \ .
\eeqa
The leading asymptotic piece in the small $a$ expansion, $G_{as}^{(0)}$,
is (up to a prefactor of $1/a^2$) equivalent to the continuum Green's function,
while the remaining terms $G_{as}^{(2i)}$, $i \ge 1$, represent
corrections of order $a^{2i}$ with respect to $G_{as}^{(0)}$ and vanish in the 
continuum limit $a \to 0$.


\section{Coefficients $\mbox{\boldmath$\kappa$}$ and $\mbox{\boldmath$\lambda$}$}  \label{app:coeff}

In this appendix, we present the coefficients $\kappa$ and $\lambda$
which appear in Eqs.~(\ref{eq:Ad}),  
(\ref{eq:Adtilde}) and~(\ref{eq:Bd}). 
The coefficients $\kappa$ in Eq.~(\ref{eq:Ad}) are given in $d$ dimensions by
\beqa
\kappa_0&=&\frac{2^{1-3 d} m^{-6+3 d} \pi ^{2-\frac{3 d}{2}}  
\Gamma\left(4-\frac{3 d}{2}\right) \sin\left(\frac{3 d \pi }{2}\right)}{
\Gamma\left(2-\frac{d}{2}\right)^3\sin\left(\frac{d \pi }{2}\right)^3},\\
\kappa_1&=&\frac{2^{1-3 d} m^{-6+3 d} \pi ^{2-\frac{3 d}{2}} (-3 \Gamma\left(5-\frac{3 d}{2}\right) \Gamma\left(2-\frac{d}{2}\right)+\Gamma\left(4-\frac{3
d}{2}\right) \Gamma\left(3-\frac{d}{2}\right)) \sin\left(\frac{3 d \pi }{2}\right)}{\Gamma\left(2-\frac{d}{2}\right)^3 \Gamma\left(3-\frac{d}{2}\right)\sin\left(\frac{d \pi }{2}\right)^3},\\
\kappa_2&=&\frac{2^{-3 d} m^{-6+3 d} \pi ^{2-\frac{3 d}{2}} (-6 \Gamma\left(5-\frac{3 d}{2}\right) \Gamma\left(2-\frac{d}{2}\right)+\Gamma\left(4-\frac{3
d}{2}\right) \Gamma\left(3-\frac{d}{2}\right)) \sin\left(\frac{3 d \pi }{2}\right)}{\Gamma\left(2-\frac{d}{2}\right)^3 \Gamma\left(3-\frac{d}{2}\right)\sin\left(\frac{d \pi }{2}\right)^3}\no\\
&+&
\frac{3\cdot 2^{-3 d} m^{-6+3 d} \pi ^{2-\frac{3 d}{2}} \Gamma\left(6-\frac{3 d}{2}\right) (\Gamma\left(3-\frac{d}{2}\right)^2+2
\Gamma\left(2-\frac{d}{2}\right) \Gamma\left(4-\frac{d}{2}\right)) \sin\left(\frac{3 d \pi }{2}\right)}{\Gamma\left(2-\frac{d}{2}\right)^2 \Gamma\left(3-\frac{d}{2}\right)^2
\Gamma\left(4-\frac{d}{2}\right)\sin\left(\frac{d \pi }{2}\right)^3},\\
\kappa_3&=&\frac{2^{-3 d} m^{-6+3 d} \pi ^{2-\frac{3 d}{2}} ((-4+d) \Gamma\left(4-\frac{3 d}{2}\right)+18 \Gamma\left(5-\frac{3
d}{2}\right)) \sin\left(\frac{3 d \pi }{2}\right)}{3 (-4+d) \Gamma\left(2-\frac{d}{2}\right)^3\sin\left(\frac{d \pi }{2}\right)^3}\no\\
&+&\frac{2^{4-3 d} (-22+3 d) m^{-6+3 d} \pi ^{2-\frac{3 d}{2}}  \Gamma\left(7-\frac{3 d}{2}\right) \sin\left(\frac{3 d \pi
}{2}\right)}{(-8+d) (-6+d) (-4+d)^2 \Gamma\left(2-\frac{d}{2}\right)^3\sin\left(\frac{d \pi }{2}\right)^3},\\
\kappa_4&=&-\frac{3\cdot 2^{5-4 d} m^{-6+3 d} \pi ^{\frac{1}{2}-\frac{3 d}{2}} \cot\left(\frac{d \pi }{2}\right) \Gamma\left(\frac{3}{2}-\frac{d}{2}\right)}{-2+d},\\
\kappa_5&=&\frac{3\cdot 2^{2-3 d} m^{-6+3 d} \pi ^{1-\frac{3 d}{2}} \cot\left(\frac{d \pi }{2}\right) ((-4+d) d \Gamma\left(3-d\right)+4 d \Gamma\left(4-d\right)+2 \Gamma\left(5-d\right))}{(-2+d)
d \Gamma\left(3-\frac{d}{2}\right)},\\
\kappa_6&=&\frac{3\cdot 2^{2-3 d} m^{-6+3 d} \pi ^{1-\frac{3 d}{2}} \cot\left(\frac{d \pi }{2}\right) ((4-d) d (2+d) \Gamma\left(3-d\right)+8 (5+d) \Gamma\left(5-d\right)) \Gamma\left(4-\frac{d}{2}\right)}{(-6+d)
d (-4+d^2) \Gamma\left(3-\frac{d}{2}\right)^2}\no\\
&+&
\frac{3\cdot 2^{4-3 d} (2+d) m^{-6+3 d} \pi ^{1-\frac{3 d}{2}} \cot\left(\frac{d \pi }{2}\right) \Gamma\left(5-d\right)}{(-6+d) (-4+d^2) \Gamma\left(3-\frac{d}{2}\right)},\\
\kappa_7&=&\frac{3\cdot 2^{1-3 d} m^{-6+3 d} \pi ^{2-\frac{3 d}{2}}}{\Gamma\left(\frac{d}{2}\right)^2\sin\left(\frac{d \pi }{2}\right)^2},\\
\kappa_8&=&\frac{3\cdot 2^{3-3 d} m^{-6+3 d} \pi ^{1-\frac{3 d}{2}} \Gamma\left(3-\frac{d}{2}\right)}{(-2+d) \Gamma\left(1+\frac{d}{2}\right)\sin\left(\frac{d \pi }{2}\right)}.
\eeqa
The coefficients $\lambda$ in Eq.~(\ref{eq:Bd}) are given by
\beqa
\lambda_0&=&0 \ ,\\
\lambda_1&=&-\frac{2^{-1-3 d} m^{-8+3 d} \pi ^{2-\frac{3 d}{2}} \Gamma\left(5-\frac{3 d}{2}\right) (d \Gamma\left(2-\frac{d}{2}\right)+2
\Gamma\left(3-\frac{d}{2}\right)) \sin\left(\frac{3 d \pi }{2}\right)}{\Gamma\left(2-\frac{d}{2}\right)^3 \Gamma\left(3-\frac{d}{2}\right)\sin\left(\frac{d \pi }{2}\right)^3},\\
\lambda_2&=&-\frac{2^{-1-3 d} m^{-8+3 d} \pi ^{2-\frac{3 d}{2}} \Gamma\left(5-\frac{3 d}{2}\right) (d \Gamma\left(2-\frac{d}{2}\right)+2
\Gamma\left(3-\frac{d}{2}\right)) \sin\left(\frac{3 d \pi }{2}\right)}{\Gamma\left(2-\frac{d}{2}\right)^3 \Gamma\left(3-\frac{d}{2}\right) \sin\left(\frac{d \pi }{2}\right)^3}\no
\eeqa
\beqa
&+&
\frac{2^{-3 d} m^{-8+3 d} \pi ^{2-\frac{3 d}{2}}  \Gamma\left(6-\frac{3 d}{2}\right) (d \Gamma\left(2-\frac{d}{2}\right)+3 \Gamma\left(3-\frac{d}{2}\right))
\sin\left(\frac{3 d \pi }{2}\right)}{\Gamma\left(2-\frac{d}{2}\right)^2 \Gamma\left(3-\frac{d}{2}\right)^2\sin\left(\frac{d \pi }{2}\right)^3}\no\\
&+&
\frac{2^{-1-3 d} d m^{-8+3 d} \pi ^{2-\frac{3 d}{2}} \Gamma\left(6-\frac{3 d}{2}\right) \sin\left(\frac{3 d \pi }{2}\right)}{\Gamma\left(2-\frac{d}{2}\right)^2
\Gamma\left(4-\frac{d}{2}\right) \sin\left(\frac{d \pi }{2}\right)^3},\\
\lambda_3&=&-\frac{2^{-2-3 d} m^{-8+3 d} \pi ^{2-\frac{3 d}{2}} (1+2 \cos\left(d \pi \right)) \Gamma\left(5-\frac{3 d}{2}\right)
(d \Gamma\left(2-\frac{d}{2}\right)+2 \Gamma\left(3-\frac{d}{2}\right))}{\Gamma\left(2-\frac{d}{2}\right)^3 \Gamma\left(3-\frac{d}{2}\right)\sin\left(\frac{d \pi }{2}\right)^2}\no\\&+&
\frac{2^{-3 d} m^{-8+3 d} \pi ^{2-\frac{3 d}{2}} (1+2 \cos\left(d \pi \right))  \Gamma\left(6-\frac{3 d}{2}\right) (d \Gamma\left(2-\frac{d}{2}\right)+3
\Gamma\left(3-\frac{d}{2}\right))}{\Gamma\left(2-\frac{d}{2}\right)^2 \Gamma\left(3-\frac{d}{2}\right)^2\sin\left(\frac{d \pi }{2}\right)^2}\no\\&-&
\frac{2^{-1-3 d} m^{-8+3 d} \pi ^{2-\frac{3 d}{2}} (1+2 \cos\left(d \pi \right))  \Gamma\left(7-\frac{3 d}{2}\right) (d \Gamma\left(2-\frac{d}{2}\right)+6
\Gamma\left(3-\frac{d}{2}\right))}{\Gamma\left(2-\frac{d}{2}\right) \Gamma\left(3-\frac{d}{2}\right)^3\sin\left(\frac{d \pi }{2}\right)^2}\no\\&+&
\frac{2^{-1-3 d} d m^{-8+3 d} \pi ^{2-\frac{3 d}{2}} (1+2 \cos\left(d \pi \right)) \Gamma\left(6-\frac{3 d}{2}\right)}{\Gamma\left(2-\frac{d}{2}\right)^2
\Gamma\left(4-\frac{d}{2}\right)\sin\left(\frac{d \pi }{2}\right)^2}\no\\&-&
\frac{3\cdot 2^{-1-3 d} m^{-8+3 d} \pi ^{2-\frac{3 d}{2}} (1+2 \cos\left(d \pi \right))  \Gamma\left(7-\frac{3 d}{2}\right) (d \Gamma\left(2-\frac{d}{2}\right)+\Gamma\left(3-\frac{d}{2}\right))}{\Gamma\left(2-\frac{d}{2}\right)^2
\Gamma\left(3-\frac{d}{2}\right) \Gamma\left(4-\frac{d}{2}\right)\sin\left(\frac{d \pi }{2}\right)^2}\no\\&-&
\frac{2^{-2-3 d} d m^{-8+3 d} \pi ^{2-\frac{3 d}{2}} (1+2 \cos\left(d \pi \right))  \Gamma\left(7-\frac{3 d}{2}\right)}{\Gamma\left(2-\frac{d}{2}\right)^2
\Gamma\left(5-\frac{d}{2}\right)\sin\left(\frac{d \pi }{2}\right)^2},\\
\lambda_4&=&-2^{2-4 d} d m^{-8+3 d} \pi ^{\frac{1}{2}-\frac{3 d}{2}} \cot\left(\frac{d \pi }{2}\right) \Gamma\left(\frac{3}{2}-\frac{d}{2}\right),\\
\lambda_5&=&\frac{2^{-1-3 d} (-4+d) d m^{-8+3 d} \pi ^{1-\frac{3 d}{2}} \cot\left(\frac{d \pi }{2}\right) \Gamma\left(3-d\right)}{\Gamma\left(3-\frac{d}{2}\right)}\no\\
&+&\frac{8^{-d}
m^{-8+3 d} \pi ^{1-\frac{3 d}{2}} \cot\left(\frac{d \pi }{2}\right) (2 d^2 \Gamma\left(4-d\right)+(6+d) \Gamma\left(5-d\right))}{(-2+d) \Gamma\left(3-\frac{d}{2}\right)},\\
\lambda_6&=&\frac{2^{-1-3 d} (-6+d) d (2+d) m^{-8+3 d} \pi ^{1-\frac{3 d}{2}} \cot\left(\frac{d \pi }{2}\right) \Gamma\left(5-d\right) \Gamma\left(3-\frac{d}{2}\right)}{(-2+d)
\Gamma\left(4-\frac{d}{2}\right)^2}\no\\
&-&
\frac{2^{-3-3 d} (-6+d) (-4+d) d m^{-8+3 d} \pi ^{1-\frac{3 d}{2}} \cot\left(\frac{d \pi }{2}\right) \Gamma\left(3-d\right)}{\Gamma\left(4-\frac{d}{2}\right)}\no\\
&+&
\frac{2^{-3 d} (-6+d) (-96+d (-12+d (3+d))) m^{-8+3 d} \pi ^{1-\frac{3 d}{2}} \cot\left(\frac{d \pi }{2}\right) \Gamma\left(5-d\right)}{(-4+d) (-2+d) d \Gamma\left(4-\frac{d}{2}\right)},\\
\lambda_7&=&\frac{2^{-3 d} d m^{-8+3 d} \pi ^{2-\frac{3 d}{2}}}{\Gamma\left(-1+\frac{d}{2}\right) \Gamma\left(\frac{d}{2}\right)\sin\left(\frac{d \pi }{2}\right)^2},\\
\lambda_8&=&\frac{2^{-3 d} (-2+(-4+d) d) m^{-8+3 d} \pi ^{2-\frac{3 d}{2}} }{\Gamma\left(\frac{d}{2}\right)^2\sin\left(\frac{d \pi }{2}\right)^2}.
\eeqa
\end{appendix}


\end{document}